 \documentclass[final,3p,11pt]{elsarticle}

\usepackage{natbib}
\biboptions{numbers,sort&compress}
\usepackage[nodots]{numcompress}
\usepackage{graphicx}
\usepackage{caption}
\usepackage{subcaption}
\usepackage{amsfonts,amssymb,amsbsy,amsmath,mathtools}

\makeatletter
\def\ps@pprintTitle{%
 \let\@oddhead\@empty
 \let\@evenhead\@empty
 \def\@oddfoot{\centerline{\thepage}}%
 \let\@evenfoot\@oddfoot}
\makeatother
\usepackage{color}
\usepackage[usenames,dvipsnames]{xcolor}
\usepackage{hyperref}

\journal{Composite Structures}
\begin{document}

\begin{frontmatter}



\renewcommand{\thefootnote}{\fnsymbol{footnote}}
\title{\textbf{Micropolar modeling approach for periodic sandwich beams\let\thefootnote\relax\footnote{{\color{Blue}\textbf{Recompiled, unedited accepted manuscript}}. \copyright 2018. Made available under \href{https://creativecommons.org/licenses/by-nc-nd/4.0/}{{\color{Blue}\textbf{\underline{CC-BY-NC-ND 4.0}}}}}}}


\author[atk,JN]{Anssi T. Karttunen\corref{cor1}}
\cortext[cor1]{Corresponding author. anssi.karttunen@iki.fi. \textbf{Cite as}: \textit{Compos. Struct.} 2018;185:656--664 \href{https://doi.org/10.1016/j.compstruct.2017.11.064}{{\color{OliveGreen}\textbf{\underline{doi link}}}}}
\address[atk]{Aalto University, Department of Mechanical Engineering, FI-00076 Aalto, Finland}
\author[JN]{J.N. Reddy}
\address[JN]{Texas A\&M University, Department of Mechanical Engineering, College Station, TX 77843-3123, USA}
\author[atk]{Jani Romanoff}

\begin{abstract}
A micropolar Timoshenko beam formulation is developed and used to model web-core sandwich beams. The beam theory is derived by a vector approach and the general solution to the governing sixth-order equations is given. A nodally-exact micropolar Timoshenko beam finite element is derived using the solution. Bending and shear stiffness coefficients for a web-core sandwich beam are determined through unit cell analysis, where the split of the shear forces into symmetric and antisymmetric parts plays a pivotal role. Static bending of web-core beams is studied using the micropolar model as well as modified couple-stress and classical Timoshenko beam models. The micropolar 1-D results are in best agreement with 2-D web-core beam frame results. This is because the micropolar beam allows antisymmetric shear deformation to emerge at locations where the 2-D web-core deformations cannot be reduced to 1-D by considering only symmetric shear behavior.
\end{abstract}

\end{frontmatter}


\section{Introduction}
Typical sandwich panels are three-layer composite structures that consist of two face sheets and a thick low-density core. The panels offer high stiffness-to-weight ratios and are widely used in transportation and construction industries. Here, we introduce a micropolar modeling approach for sandwich beams and demonstrate its robustness for modeling web-core panels that have applications in ship structures and residential buildings \cite{kujala2005,briscoe2011}.

A simple and computationally efficient way to determine the global response of a three-layer sandwich construction is to treat it as a statically equivalent single layer (ESL) beam or plate based on the first-order shear deformation theory (FSDT) \cite{reddy2004}. The use of an ESL-FSDT model requires us to determine the average bending and shear stiffness coefficients of the three-layer sandwich panel at hand. In this study, we are particularly interested in sandwich panels with unidirectional structural cores (e.g. web-, corrugated-, C-, Z-cores) that have low transverse shear stiffness coefficients along the direction perpendicular to the core. The methods for determining the conventional stiffness parameters for such panels are well-established \cite{libove1951,nordstrand1994,fung1994,fung1996,lok2000,romanoff2007,martinez2007,boorle2016,arunkumar2016,yu2017,nilsson2017}.

The ESL-FSDT approach based on classical elasticity is limited to sandwich panels with relatively thin face sheets. To elaborate on this limitation, let us consider a sandwich panel modeled using a classical ESL Timoshenko beam. The bending stiffness for the beam is calculated by applying the parallel axis theorem to the sandwich panel so that the stiffness due to the membrane action of the face sheets and the stiffness related to the local bending of the sheets with respect to their own centroid axes are summed. However, this summing procedure omits the fact that there is no such constituent in the classical Timoshenko beam theory that could account for the local bending of the face sheets. This inconsistency can be resolved by using a couple-stress Timoshenko beam model that includes a non-classical couple-stress moment \cite{ma2008,reddy2011,asghari2010,asghari2011}, which can be associated with the local bending of the faces  \cite{romanoff2014,romanoff2015,goncalves2017} (see Fig.~1). With the local bending properly accounted for, the \textit{thick-face effect} \cite{allen1969} is included in the couple-stress ESL model. Unlike the conventional ESL Timoshenko beam, the couple-stress beam has been found to give good results also for thick-faced sandwich beams \cite{goncalves2017}.
\begin{figure}
\centering
\includegraphics[scale=1]{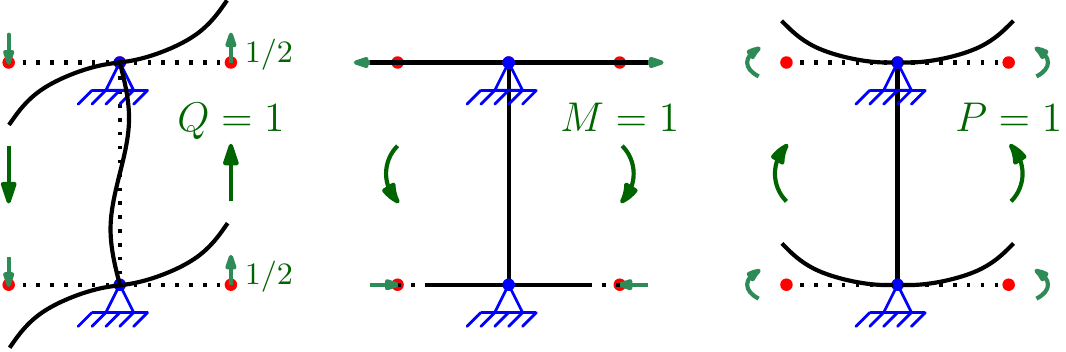}
\caption{Schematic determination of ESL web-core stiffness parameters. Stress resultants $Q=1, M=1$ and couple-stress resultant $P=1$ are divided into suitable point loads at the unit cell corners. After the corner displacements have been solved, three stiffness parameters are obtained from the resultant equations; $M$ accounts for the membrane action of the face sheets and $P$ for the local bending of the faces (i.e. thick-face effect).}
\end{figure}

The couple-stress ESL approach is an improvement to the conventional one, however, its limitations are yet to be studied. After all, couple-stress continuum theories \cite{tiersten1974,yang2002} are simplified versions of the micropolar theory \cite{eringen2012}, which may also be called, rather interchangeably, the Cosserat theory \cite{cosserat1909} or the theory of asymmetric elasticity \cite{nowacki1986}. The micropolar theory includes a microrotation which is independent of the macrorotation obtained from the displacement gradient. In other words, the microrotation is independent of the translational displacements. Couple-stress theories are arrived at through the simplifying assumption that the microrotation coincides with the macrorotation. In this work, we relax this assumption, i.e., we use a micropolar ESL Timoshenko beam to study sandwich beams. We show that a couple-stress Timoshenko beam may provide too stiff results for sandwich beams due to the inherent rotational constraint. We note that micropolar Timoshenko beam theories have been developed by several authors in recent years \cite{ramezani2009,nobili2015,regueiro2015,shaw2016,ding2016,zozulya2017}. In light of this, the main novel features of the current study are that we derive the explicit general solution to the equilibrium equations of the micropolar Timoshenko beam; use the solution to develop a nodally-exact micropolar Timoshenko beam finite element (FE) and, finally, we apply the beam model and the finite elements to practical sandwich beam problems with the micropolar ESL stiffness parameters determined through the unit cell analysis of a web-core sandwich beam.

The rest of the paper is organized as follows. In Section 2, we develop a micropolar Timoshenko beam model using a vector approach and derive the general solution to the equilibrium equations of the beam. The boundary conditions are determined from the work done by the beam stresses at the beam ends. In Section 3, a nodally-exact micropolar beam element based on the general solution is formulated. Section 4 presents the derivation of the micropolar ESL Timoshenko beam stiffness parameters through the unit cell analysis of a web-core sandwich beam. Numerical bending examples are studied in Section 5 using the classical, couple-stress and micropolar ESL Timoshenko beam theories and 2-D FE beam frame models. Finally, conclusions are drawn in Section 6.
\section{Micropolar Timoshenko beam theory}
\subsection{Two-dimensional equilibrium equations}
In addition to having independent rotational degrees of freedom, a micropolar solid can transmit couple-stresses, as well as the usual force-stresses. Fig.~2(a) shows the components of stress acting on a planar element in a varying stress field. With the body forces and couples omitted, the force and moment equilibrium of the planar element provide the stress equilibrium equations
\begin{align}
\frac{\partial \sigma_x}{\partial x}+\frac{\partial \tau_{yx}}{\partial y}&=0, \\
\frac{\partial \sigma_y}{\partial y}+\frac{\partial \tau_{xy}}{\partial x}&=0, \\
\frac{\partial m_{xz}}{\partial x}+\frac{\partial m_{yz}}{\partial y}+\tau_{xy}-\tau_{yx}&=0.
\end{align}
Note that unlike in the modified couple-stress theory \cite{lam2003}, an additional equilibrium equation for the moment of couples does not appear in the micropolar theory. We see from Eq.~(3) that the shear stresses are not necessarily symmetric (i.e., $\tau_{xy}\neq\tau_{yx}$). Further, the force-stress and couple-stress tensors are generally not symmetric in the micropolar theory. The shear stresses can be split into symmetric and antisymmetric parts \cite{mindlin1963}
\begin{align}
\tau_s=\frac{\tau_{xy}+\tau_{yx}}{2}, \\
\tau_a=\frac{\tau_{xy}-\tau_{yx}}{2},
\end{align}
respectively, as shown in Fig.~2(b). The symmetric part produces the usual shear deformation, whereas the antisymmetric part creates a rotation that causes an antisymmetric shear strain which is defined by the difference between the macrorotation and microrotation (Section 2.3). The split of the shear behavior into symmetric and antisymmetric parts greatly facilitates the determination of the equivalent sandwich stiffness parameters in Section 4.
\begin{figure}
\centering
\includegraphics[scale=1]{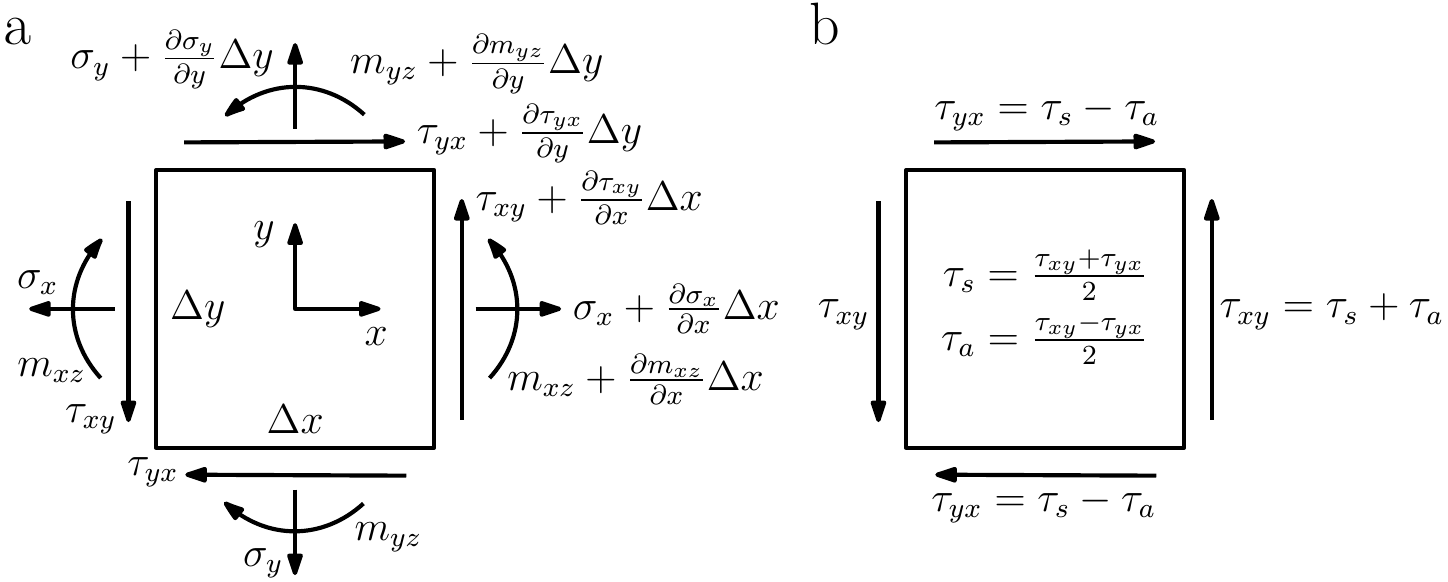}
\caption{(a) Stresses acting on a planar micropolar solid in a varying stress field. (b) Split of the shear stresses into symmetric and antisymmetric parts.}
\end{figure}
\subsection{General beam equilibrium equations}
Let us consider a beam of constant height $h$ and width $b$. In order to reduce the 2-D equilibrium equations (1)--(3) into 1-D beam equations given in terms of shear forces and moments, we multiply Eq.~(1) with $y$ and then integrate Eqs.~(1)--(3) over the cross section to obtain
\begin{align}
\frac{\partial M_x}{\partial x}-Q_{yx}&=-t, \\
\frac{\partial Q_{xy}}{\partial x}&=-q, \\
\frac{\partial P_{xz}}{\partial x}+Q_{xy}-Q_{yx}&=-m,
\end{align}
where the stress resultants are defined as
\begin{align}
M_x&=\int_A y\sigma_x\ dA,\quad P_{xz}=\int_A m_{xz}\ dA, \\
Q_{yx}&=\int_A \tau_{yx}\ dA, \quad Q_{xy}=\int_A \tau_{xy}\ dA
\end{align}
and the boundary terms resulting from integration by parts read
\begin{align}
t&=(bh/2)\left[\sigma_{yx}(x,h/2)+\sigma_{yx}(x,-h/2)\right], \\
q&=b\left[\sigma_{y}(x,h/2)-\sigma_{y}(x,-h/2)\right], \\
m&=b\left[m_{yz}(x,h/2)-m_{yz}(x,-h/2)\right].
\end{align}
The surface loads $t$ and $m$ will not be considered further in the following sections. Only the pressure load $q$ is of practical interest to us.
\subsection{Displacements, strains and stresses of Timoshenko beam}
We now consider the micropolar Timoshenko beam presented in Fig.~3. The length of the beam is $L$ and, in line with the foregoing, the beam has a rectangular cross-section of constant width $b$ and height $h$. The kinematic description of the beam is assumed to take the form
\begin{equation}
U_x(x,y)=y\phi(x), \quad U_y(x,y)=u_y(x), \quad \Psi(x,y)=\psi(x),
\end{equation}
where $\phi$ is the rotation of the cross-section at the central axis of the beam, $u_y$ is the transverse deflection and $\psi$ is an independent microrotation.
\begin{figure}[hb]
\centering
\includegraphics[scale=0.68]{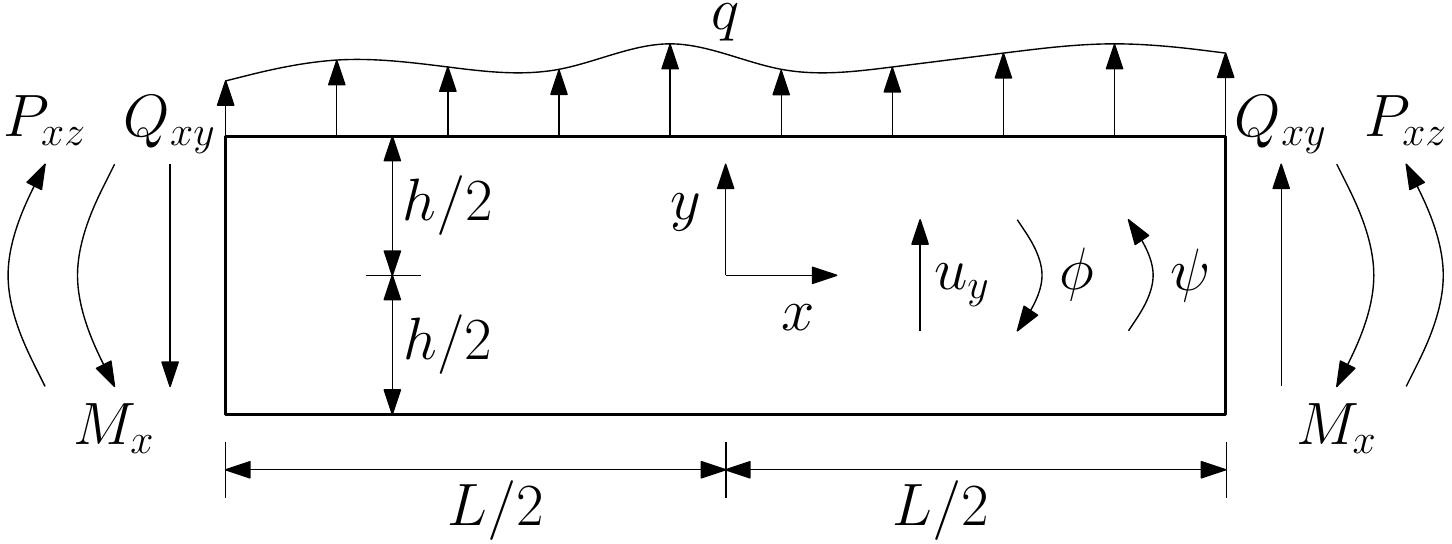}
\caption{Micropolar Timoshenko beam. The positive directions of the stress resultants and displacements are shown.}
\end{figure}

The axial normal strains $\epsilon_x$ and $\epsilon_y$, and the relative strains $\epsilon_{xy}$ and $\epsilon_{yx}$ of the beam are \cite{borst1991}
\begin{equation}
\begin{aligned}
\epsilon_x&=\frac{\partial U_x}{\partial x}=y\phi', \ & \epsilon_{xy}&=\frac{\partial U_y}{\partial x}-\Psi=u_y'-\psi, \\ \epsilon_y&=\frac{\partial U_y}{\partial y}=0, \ & \epsilon_{yx}&=\frac{\partial U_x}{\partial y}+\Psi=\phi+\psi,
\end{aligned}
\end{equation}
where the prime ``$'$" on the variables denotes differentiation with respect to $x$. The components of the relative strains are illustrated in Fig.~4. The microstructure of the planar element exhibits a rigid microrotation $\Psi$. The rotating axes shown for one material point in the microstructure are called rigid orthogonal directors in the micropolar theory. In the micromorphic theory, the directors of each material point are deformable \cite{eringen2012}.
\begin{figure}
\centering
\includegraphics[scale=1.1]{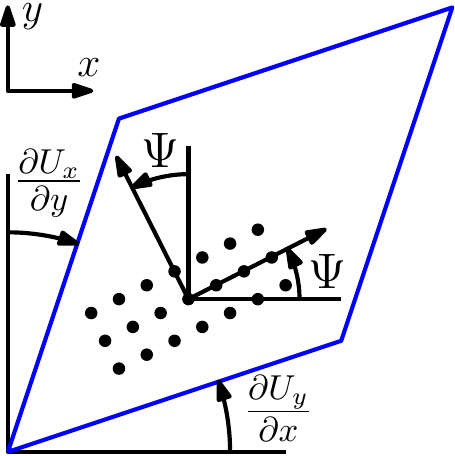}
\caption{Components of relative strains $\epsilon_{xy}$ and $\epsilon_{yx}$. The rigid rotation of the microstructure is described by $\Psi$.}
\end{figure}
The symmetric and antisymmetric shear strains are defined as
\begin{align}
\gamma_s&=\epsilon_{xy}+\epsilon_{yx}=u_y'+\phi, \\ \gamma_a&=\epsilon_{xy}-\epsilon_{yx}=u_y'-\phi-2\psi,
\end{align}
respectively. We can see that the symmetric part takes the same form as the shear strain in the classical Timoshenko beam theory. The antisymmetric part is twice the difference between the usual macrorotation and the microrotation. The curvatures that are energetically conjugate to the couple stresses are
\begin{equation}
\kappa_{xz}=\frac{\partial \Psi}{\partial x}=\psi', \quad \kappa_{yz}=\frac{\partial \Psi}{\partial y}=0.
\end{equation}
The curvatures describe the bending of the planar element (Fig.~2) due to the couple-stresses. In the case of a sandwich beam, $\kappa_{xz}$ represents the curvature of the face sheets bent by the local moments (see Fig.~1).

For the 1-D micropolar Timoshenko beam, the isotropic stress-strain relations can be written as \cite{borst1991}
\begin{equation}
\begin{Bmatrix}
\sigma_x \\
\tau_{xy} \\
\tau_{yx} \\
m_{xz}
\end{Bmatrix}=
\begin{bmatrix}
E & 0 & 0 & 0 \\
0 & G+G_c & G-G_c & 0 \\
0 & G-G_c & G+G_c & 0 \\
0 & 0 & 0 & 2Gl^2
\end{bmatrix}
\begin{Bmatrix}
\epsilon_x \\
\epsilon_{xy} \\
\epsilon_{yx} \\
\kappa_{xz}
\end{Bmatrix},
\end{equation}
where $E$ and $G$ are the Young's modulus and shear modulus, respectively, and $G_c$ and $l$ are the Cosserat modulus and microlength, respectively.
\subsection{Timoshenko beam equations}
By using the constitutive relations (19), Eqs.~(9) and (10) yield the stress resultants
\begin{align}
M_x&=D_x\phi', \\ P_{xz}&=2D_{xz}\kappa_{xz}=2D_{xz}\psi', \\
Q_{xy}&=D_s(u_y'+\phi)+D_a(u_y'-\phi-2\psi), \\
Q_{yx}&=D_s(u_y'+\phi)-D_a(u_y'-\phi-2\psi).
\end{align}
Moreover, the symmetric and antisymmetric shear forces are defined as
\begin{align}
Q_s&=\frac{Q_{xy}+Q_{yx}}{2}=D_s(u_y'+\phi), \\
Q_a&=\frac{Q_{xy}-Q_{yx}}{2}=D_a(u_y'-\phi-2\psi),
\end{align}
respectively. For an isotropic, homogeneous material we have
\begin{equation}
D_x=EI,\ D_{xz}=GAl^2,\ D_s=GA,\ D_a=G_cA.
\end{equation}
For ESL beams, the stiffness parameters $D_x$, $D_{xz}$, $D_s$ and $D_a$ are determined in an average sense from a unit cell. This procedure is carried out for a web-core beam in Section 4. The stiffness parameters can also be taken to represent, for example, a functionally graded material \cite{reddy2011}.

By substituting Eqs.~(20)--(23) into the equilibrium equations (6)--(8), we arrive at ($t=m=0$)
\begin{align}
D_x\phi''-D_s(u_y'+\phi)+D_a(u_y'-\phi-2\psi)&=0, \\
D_s(u_y''+\phi')+D_a(u_y''-\phi'-2\psi')&=-q, \\
2D_{xz}\psi''+2D_a(u_y'-\phi-2\psi)&=0.
\end{align}
Furthermore, we can write
\begin{equation}
M_x''-P_{xz}''=-q \quad \rightarrow \quad \psi'''=\frac{1}{2D_{xz}}(D_x\phi'''+q).
\end{equation}
The closed-form solution to Eqs.~(27)--(29) can be obtained by first decoupling the equations. The solution process is explained in Appendix A. Alternatively, Eqs.~(27)--(29) can be readily solved in the given form by a symbolic mathematical tool such as Maple. The homogeneous solution ($q=0$) reads
\begin{align}
u_y&=c_1+c_2x+\frac{1}{2}c_3x^2-c_4\left[\left(\frac{D_x+D_{xz}}{D_s}+\frac{D_{xz}}{D_a}\right)x-\frac{x^3}{3}\right]+\alpha\left(c_5\textrm{e}^{\beta x}-c_6\textrm{e}^{-\beta x}\right), \\
\phi&=-c_2-c_3x-c_4\left(\frac{D_x+D_{xz}}{D_s}-\frac{D_{xz}}{D_a}+x^2\right)+\frac{2D_{xz}}{D_x}\left(c_5\textrm{e}^{\beta x}+c_6\textrm{e}^{-\beta x}\right), \\
\psi&=c_2+c_3x+c_4x^2+c_5\textrm{e}^{\beta x}+c_6\textrm{e}^{-\beta x} ,
\end{align}
where
\begin{equation}
\alpha^2=\frac{2D_{xz}[\left(D_x+D_{xz}\right)D_a-D_sD_{xz}]^2}{D_xD_sD_a(D_x+2D_{xz})(D_s+D_a)}, \ \beta^2=\frac{2D_sD_a(D_x+2D_{xz})}{D_xD_{xz}(D_s+D_a)}.
\end{equation}
As an example, the particular solution to be added to the homogeneous solution in the case of a uniformly distributed load $q(x)=q_0$ is
\begin{align}
q_{uy}&=-\frac{q_0x^2[6D_xD_a+6D_{xz}(D_s+D_a)-D_sD_ax^2]}{24D_s(D_x+2D_{xz})D_a}, \\
q_{\phi}&=-\frac{q_0x[3D_a(D_x+D_{xz})-D_s(3D_{xz}-D_ax^2)]}{6D_sD_a(D_x+2D_{xz})}, \\
q_{\psi}&=\frac{q_0x^3}{6(D_x+2D_{xz})}.
\end{align}
For example, Eq.~(35) is added to the RHS of Eq.~(31).
\subsection{Clapeyron's theorem and boundary conditions}
To obtain the boundary conditions for the beam, let us consider the strain energy of the beam and the work done by surface tractions. The strain energy is
\begin{equation}
\begin{aligned}
U&=\frac{1}{2}\int_V(\sigma_x\epsilon_x+\tau_{xy}\epsilon_{xy}+\tau_{yx}\epsilon_{yx}+m_{xz}\kappa_{xz})dV \\
&=\frac{1}{2}\int_{-L/2}^{L/2}\left[M_x\phi'+Q_{xy}(u_y'-\psi)+Q_{yx}(\phi+\psi)+P_{xz}\psi'\right]dx.
\end{aligned}
\end{equation}
An alternative form for the strain energy may be attained by noting that
\begin{equation}
\tau_{xy}\epsilon_{xy}+\tau_{yx}\epsilon_{yx}=\tau_{s}\gamma_{s}+\tau_{a}\gamma_{a}.
\end{equation}
The stresses calculated from Eq.~(19) exist also on the lateral end surfaces of the beam where they act as surface tractions and bring about the work
\begin{equation}
\begin{aligned}
W_s&=\int_A\left[(\sigma_xU_x+\tau_{xy}U_y+m_{xz}\Psi)|_{L/2}-(\sigma_xU_x+\tau_{xy}U_y+m_{xz}\Psi)|_{-L/2}\right]dA \\
&=\left[M_x\phi+Q_{xy}u_y+P_{xz}\psi\right]_{-L/2}^{L/2}.
\end{aligned}
\end{equation}
The work due to a distributed pressure load is
\begin{equation}
W_q=\int_{-L/2}^{L/2}qu_y\ dx.
\end{equation}
According to Clapeyron's theorem, the strain energy stored in a linear elastic body is equal to one-half of the work done by the surface tractions and body forces if they were to move (slowly) through their respective displacements from an unstressed state to the state of equilibrium \cite{sadd2014, reddy2017}. In the present micropolar case, Clapeyron's theorem leads to
\begin{equation}
2U-W_s-W_q=0,
\end{equation}
which is verified by the general solution (31)--(37) for $q=q_0$ by the aid of Mathematica. The boundary conditions are taken from the conjugate pairs of Eq.~(40) for both ends of the beam as
\begin{equation}
M_x \quad \textrm{or} \quad \phi, \qquad  Q_{xy} \quad \textrm{or} \quad u_y, \qquad  P_{xz} \quad \textrm{or} \quad \psi.
\end{equation}
If we were to derive the weak form for the beam by starting from Eqs.\ (6)--(8), the same boundary conditions would be obtained. Note that a variational formulation for the micropolar Timoshenko beam can be carried out by taking the first variation of Eq.~(42). For details, see Refs.\ \cite{karttunen2017,reddy2017}.
\section{Micropolar Timoshenko beam element}
\begin{figure}
\centering
\includegraphics[scale=0.65]{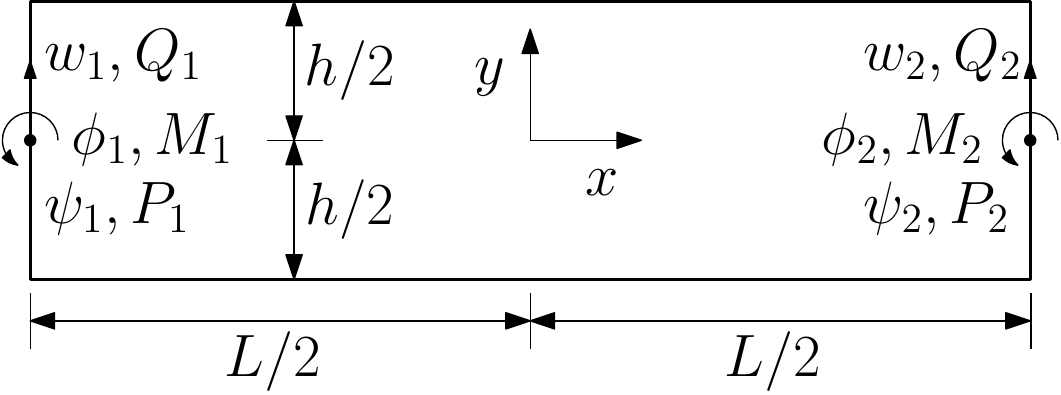}
\caption{Set-up according to which the micropolar Timoshenko beam finite element is developed.}
\end{figure}
The general displacement solution (31)--(33) is used as the basis for the derivation of a nodally-exact beam finite element. A concise formulation is given below and a more detailed one with lengthy explicit expressions is given in the supplementary Mathematica file \textit{MicrobeamFE} provided online. Fig.~5 presents the setting according to which the finite element is developed. Both nodes have three degrees of freedom, namely, transverse displacement $w_{i}$ and rotations $\phi_{i}$ and $\psi_{i}$ ($i=1,2$). Using Eqs.~(31)--(33), we define the FE degrees of freedom as
\begin{equation}
\begin{aligned}
        w_{1}&=u_y(-L/2), & \qquad w_{2}&=u_y(L/2), \\
        \phi_1&=-\phi(-L/2), & \qquad \phi_2&=-\phi(L/2) \\
        \psi_1&=\psi(-L/2), & \qquad \psi_2&=\psi(L/2).
\end{aligned}
\end{equation}
In matrix form we have
\begin{equation}
\mathbf{u}=\mathbf{H}\mathbf{c},
\end{equation}
where $\mathbf{u}$ is the displacement vector, $\mathbf{H}$ is a coefficient matrix and $\mathbf{c}$ contains the constant coefficients $c_i$ ($i=1,\ldots,6$), which are obtained in terms of the FE degrees of freedom by
\begin{equation}
\mathbf{c}=\mathbf{H}^{-1}\mathbf{u}.
\end{equation}
The kinematic variables (31)--(33) in terms of the FE degrees of freedom may then be written as
\begin{align}
\begin{Bmatrix}
u_y(x) \\
\phi(x)\\
\psi(x)
\end{Bmatrix}
=\mathbf{A}\mathbf{c}=\mathbf{A}\mathbf{H}^{-1}\mathbf{u}=
\begin{Bmatrix}
\mathbf{N}_{uy} \\
\mathbf{N}_{\phi} \\
\mathbf{N}_{\psi}
\end{Bmatrix}\mathbf{u},
\end{align}
where $\mathbf{A}$ is a matrix with polynomial and exponential terms and $\mathbf{N}_{uy}$, $\mathbf{N}_{\phi}$ and $\mathbf{N}_{\psi}$ contain the shape functions. The nodal forces for the finite element are
\begin{equation}
\begin{aligned}
        Q_1&=-Q_{xy}(-L/2) , & \qquad Q_2&=Q_{xy}(L/2), \\
        M_1&=M_x(-L/2), & \qquad M_2&=-M_x(L/2), \\
        P_1&=-P_{xz}(-L/2), & \qquad P_2&=P_{xz}(L/2).
\end{aligned}
\end{equation}
In matrix form we have
\begin{equation}
\mathbf{f}=\mathbf{G}\mathbf{c},
\end{equation}
where $\mathbf{f}$ is the nodal load vector and $\mathbf{G}$ is a coefficient matrix. Using Eqs.~(46) and (49) we obtain
\begin{equation}
\mathbf{K}\mathbf{u}=\mathbf{f},
\end{equation}
where the stiffness matrix is
\begin{equation}
\mathbf{K}=\mathbf{G}\mathbf{H}^{-1}.
\end{equation}
Once the nodal displacements $\mathbf{u}$ have been solved from (50) for a set of boundary conditions and nodal loads, they are substituted into Eq.~(47) to obtain the continuous central axis displacements.
\section{Application to a periodic web-core beam}
The micropolar Timoshenko beam model is taken to represent a periodic web-core sandwich beam and we determine the ESL stiffness parameters $D_x$, $D_{xz}$ and $D_s$ by a force approach from the unit cell shown in Fig.~6. Antisymmetric shear stiffness $D_a$ is calculated using a strain energy approach. A half of the unit cell is considered to represent an infinitesimal length $s\approx dx$. The distance between the centerlines of the faces, that is, the height of the unit cell is $h$. Bending rigidities of the face and web plates are $EI_f$ and $EI_w$, respectively, and $EA_f$ and $EA_w$ are the axial rigidities. The unit cell is modeled using five Euler--Bernoulli beam finite elements.
\begin{figure}[ht]
\centering
\includegraphics[scale=0.9]{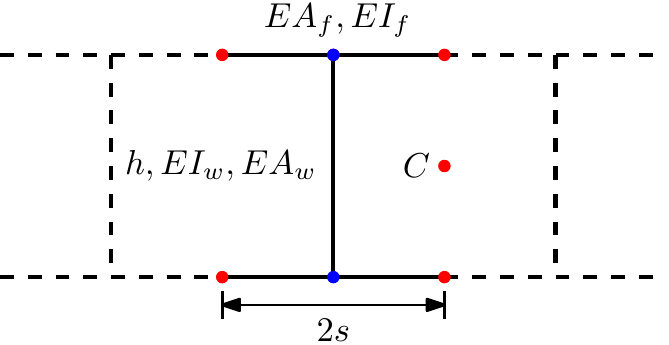}
\caption{Infinitesimal unit cell of a web-core beam.}
\end{figure}

Three unit resultant load cases are studied in Fig.~7. Each resultant load on the side of a unit cell is considered to be acting initially at point $C$ which represents the midpoint of an undeformed microstructural building block. For analysis, the unit loads are divided into equivalent point loads acting in the unit cell corner nodes. In the first case ($Q_{s,C} = 1$), axial forces of magnitude $s/h$ have been exerted on the faces to keep the unit cell in moment equilibrium before the boundary conditions are applied \cite{lok2000}. We also note that due to symmetric geometry and loading of the unit cell, the transverse inextensibility constraint $u_{y}^A = u_{y}^B$ between the faces is satisfied. Moreover, the stiffness parameters $D_s$ and $D_a$ fully account for the shear behavior of the unit cell and additional shear correction factors are not needed. If the constituents of the unit cell are thick, they can be modeled using classical Timoshenko beam elements, in which case a shear correction factor would be included in the stiffness parameters.
\begin{figure}[hb]
\centering
\includegraphics[scale=0.99]{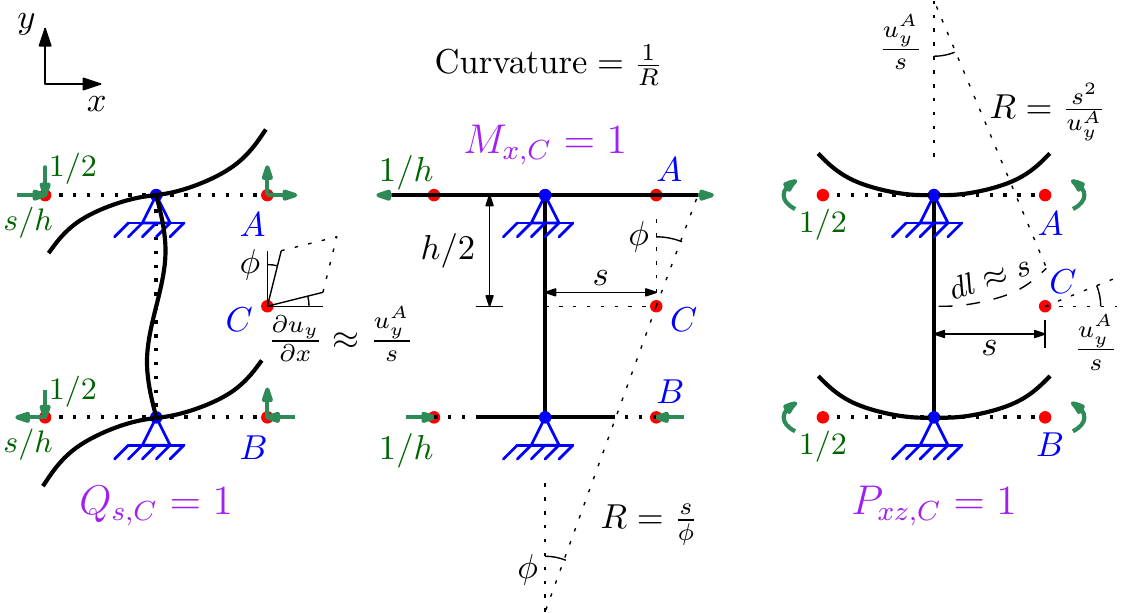}
\caption{Three unit resultant load cases that constitute the behavior of a web-core beam under bending. The analysis of each case results in kinematic variables that have been averaged over the unit cell.}
\end{figure}

We have chosen point $C$ in Fig.~7 to be of particular interest because the unit cell is considered to be under uniform stress and strain conditions, that is, it represents a point. Then the rotations need to be constant over the unit cell for them to be representative of, for example, the symmetric shear deformation (see point $C$ in the first case of Fig.~7). This amounts to treating any face slope as constant. Rotation $\phi$ is determined from the axial displacements of points $A$ and $B$ in Fig.~7.

In summary, by calculating the nodal displacements of the finite element based unit cell in each load case, we can determine averaged central axis kinematic variables that relate to the symmetric shear strain and curvature $1/R$ of the unit cell. To determine the stiffness parameters $D_s$, $D_x$ and $D_{xz}$ the average variables are used in the unit stress resultant equations $Q_{s,C}=1$, $M_{x,C}=1$ and $P_{xz,C}=1$, respectively [cf.~Eqs.~(24), (20) and (21)].

For the determination of shear stiffness $D_s$ by the unit load method according to the first case in Fig.~7, we have the rotations
\begin{equation}
\phi=\frac{2s^2}{h^2EA_f} \quad \textrm{and} \quad \frac{\partial u_y}{\partial x}\approx\frac{u_y^A}{s}=\frac{s}{6}\left(\frac{h}{EI_w}+\frac{s}{EI_f}\right).
\end{equation}
It follows that for $Q_{s,C}=1$ we obtain from Eq.~(24)
\begin{equation}
D_s=\frac{6}{s}\left(\frac{12s}{h^2EA_f}+\frac{s}{EI_f}+\frac{h}{EI_w}\right)^{-1} \ [\textrm{N}].
\end{equation}
In the bending cases $M_{x,C}=1$ and $P_{xz,C}=1$, the curvatures $1/R$ of the unit cell lead to
\begin{align}
D_x=\frac{h^2EA_f}{2} \ [\textrm{Nm}^2] \quad \textrm{and} \quad D_{xz}=2EI_f \ [\textrm{Nm}^2]
\end{align}
through Eqs.~(20) and (21), respectively. We see that $D_{xz}$ is the sum of the local bending stiffnesses of the two faces and within the unit cell we have $\psi\approx u_y^A/s$.

The symmetric shear stiffness $D_s$ was basically determined from a case in which we had $Q_{s,C}=1$ and $Q_{a,C}=0$ on the sides of the unit cell. In the configuration of Fig.~7, the opposite case $Q_{s,C}=0$ and $Q_{a,C}=1$ leads to the same nodal displacements and overall deformation [see, e.g., Eq.~(52)] because the symmetric and antisymmetric shear forces have the same directions on the lateral sides of the unit cell (see Fig.~2). Under a unit load $Q_{a,C}=1$, the strain energy of the unit cell is
\begin{equation}
U_{\textrm{cell}}=\frac{1}{2}\left(\mathbf{u}^{\textrm{T}}\mathbf{K}\mathbf{u}\right)_{\textrm{cell}}.
\end{equation}
Next, we assume that the unit cell is identical to a short micropolar beam of length $L=2s$. Within the unit cell we have, again, $\psi\approx u_y^A/s$ so that on the basis of Eqs.~(25), (38), and (39) the strain energy of the beam due to pure antisymmetric shear loading $Q_a=1$ becomes
\begin{equation}
U_{\textrm{beam}}\approx sD_a\left(-\frac{u_y^A}{s}-\phi\right)^2,
\end{equation}
where $u_y^A/s$ and $\phi$ are the same as given by Eq.~(52). The strain energies of the unit cell and the beam are equal under identical deformations
\begin{equation}
U_{\textrm{beam}}=U_{\textrm{cell}}
\end{equation}
we obtain
\begin{equation}
D_a=\frac{6}{s}\left(\frac{12s}{h^2EA_f}+\frac{s}{EI_f}+\frac{h}{EI_w}\right)^{-1} \ [\textrm{N}].
\end{equation}
We have $D_a=D_s$.
\section{Numerical case studies}
\subsection{General setup}
In this section we study the three beam calculation examples presented in Figs.~8(a), 8(b) and 8(c). The beams consist of rectangular web-core blocks (cf. Fig.~6) and are similar to those tested for static bending in \cite{karttunen2017b}. The face sheets and webs are made of steel ($E=210$ GPa, $\nu=0.3$). The joints between the faces and webs are modeled here as rigid. The beam width and height are $b=0.05$ m and $h=0.043$ m, respectively, in all cases. The web spacing is $2s=0.12$ m and the face and web thicknesses are 3 mm and 4 mm, respectively. In addition to the micropolar ESL beam, we calculate the bending responses using classical and modified couple-stress ESL Timoshenko beams for which the shear stiffness $D_Q=D_s$ is given by Eq.~(53) and the bending stiffness $D_x$ by Eq.~(54); for the couple-stress beam we also have $S_{xy}=8EI_f=4D_{xz}$ \cite{goncalves2017}. Reference solutions are calculated using 2-D FE Euler--Bernoulli beam frames modeled by Abaqus; the pins in simply-supported cases are at the central axis of the frame so that the model corresponds to 1-D cases.
\begin{figure}
\centering
\includegraphics[scale=0.74]{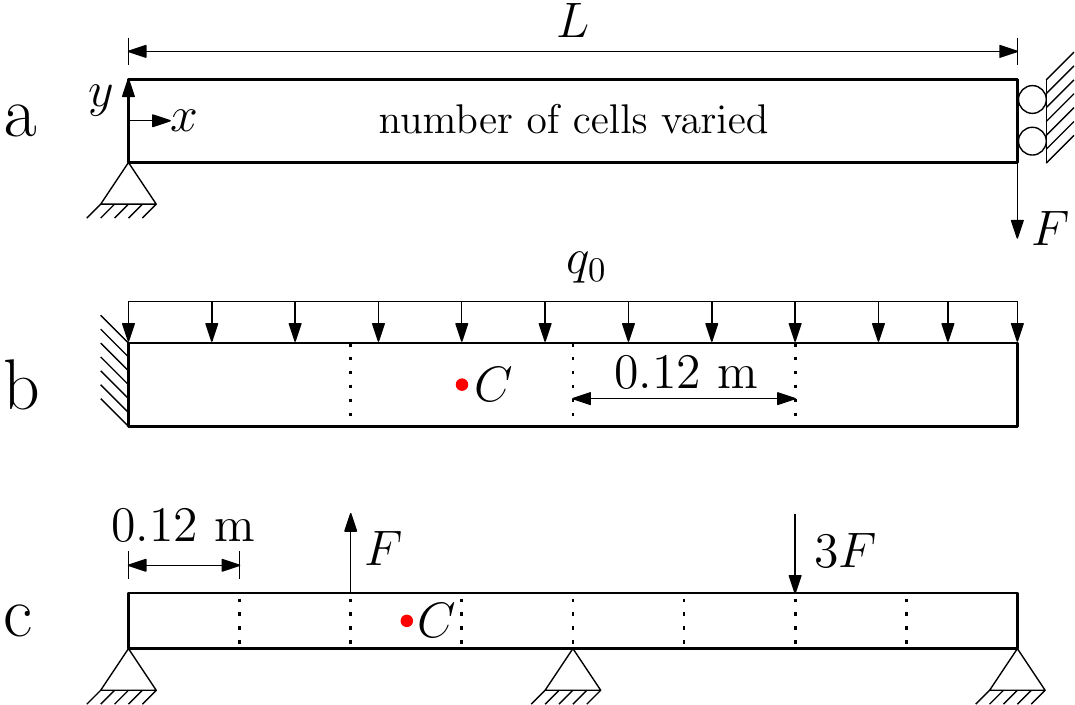}
\caption{(a) Three-point bending of a web-core sandwich beam modeled by a symmetric half. (b) Web-core cantilever beam under a uniformly distributed load. (c) Web-core beam on three supports modeled by using four beam elements.}
\end{figure}
\subsection{Numerical results and discussion}
The boundary conditions for the three-point bending case in Fig.~8(a) that are used to solve the integration constants in Eqs.~(31)--(33) are
\begin{equation}
\begin{aligned}
x&=0:\ u_y=0, \ M_x=P_{xz}=0 , \\ x&=L:\ \phi=\psi=0, \ Q_{xy}=-F,
\end{aligned}
\end{equation}
where the point load is now $F=500$ N. Figure~9(a) shows the transverse deflections of the different ESL Timoshenko beam models and Fig.~9(b) shows the errors calculated from
\begin{equation}
\Delta u_y=100\times\frac{u_y^{\textrm{1-D Timoshenko}}-u_y^{\textrm{2-D FE}}}{u_y^\textrm{2-D FE}} \ [\%]
\end{equation}
in terms of the maximum deflection. The classical and couple-stress modeling approaches result in considerably larger errors than the micropolar approach for short beams. As the beams become longer, the errors become smaller. To understand the reason for this, we look at the rotation variables of the micropolar beam displayed in Fig.~10(a) and the shear forces calculated from the rotations in Fig~10(b). The solid line in Fig.~10(a) shows that the difference between the macrorotation and microrotation is non-zero only near the slider support. The difference is directly proportional to the antisymmetric shear strain $\gamma_a$ and shear force $Q_a$. Figure 10(b) includes a schematic depiction with linearized face slopes on how the web-core blocks along the beam deform according to the 2-D FE beam frame results. We can see that when located at sufficient distance from the slider support, a web-core block exhibits only symmetric shear strain ($\gamma_a=0$) in terms of the 1-D micropolar model. The antisymmetric shear strain appears due to the slider support which causes the 2-D web-core block next to it to deform in a manner which cannot be described only by symmetric shear strain in 1-D. Only the micropolar approach can capture the antisymmetric behavior. Nevertheless, since the antisymmetric behavior occurs only in the immediate vicinity of the slider support, the results given by the classical and couple-stress ESL beams improve overall as the beam becomes longer and the number of the blocks increases as shown in Fig.~9(b).
\begin{figure}
\centering
\includegraphics[scale=0.48]{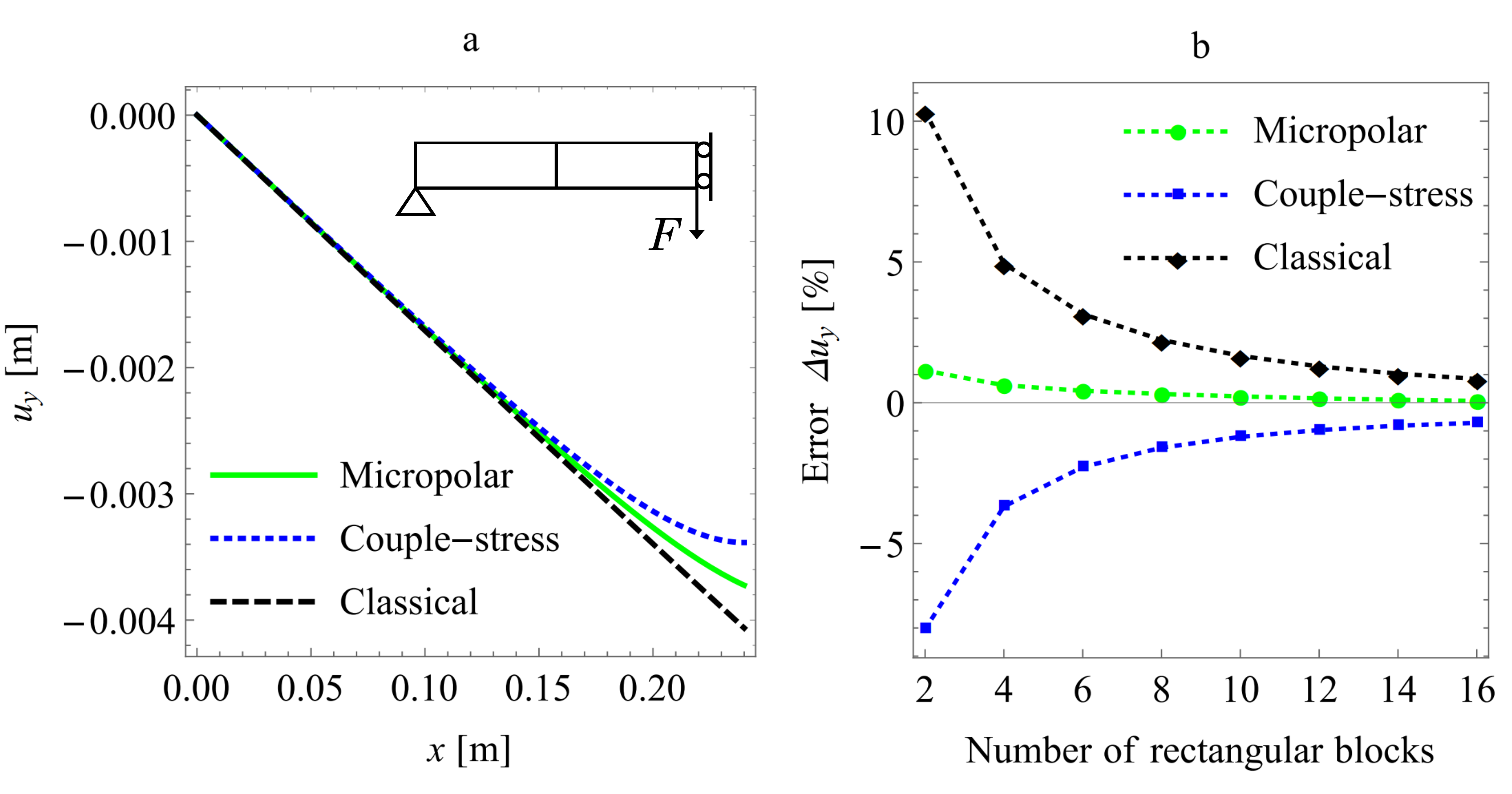}
\caption{(a) Transverse deflections of 1-D classical, couple-stress and micropolar ESL Timoshenko beams under three-point bending. (b) Errors of the 1-D beam models in terms of maximum deflection in comparison to 2-D FE beam frame solution (face sheet deflection) calculated using Abaqus.}
\end{figure}
\begin{figure}
\centering
\includegraphics[scale=0.44]{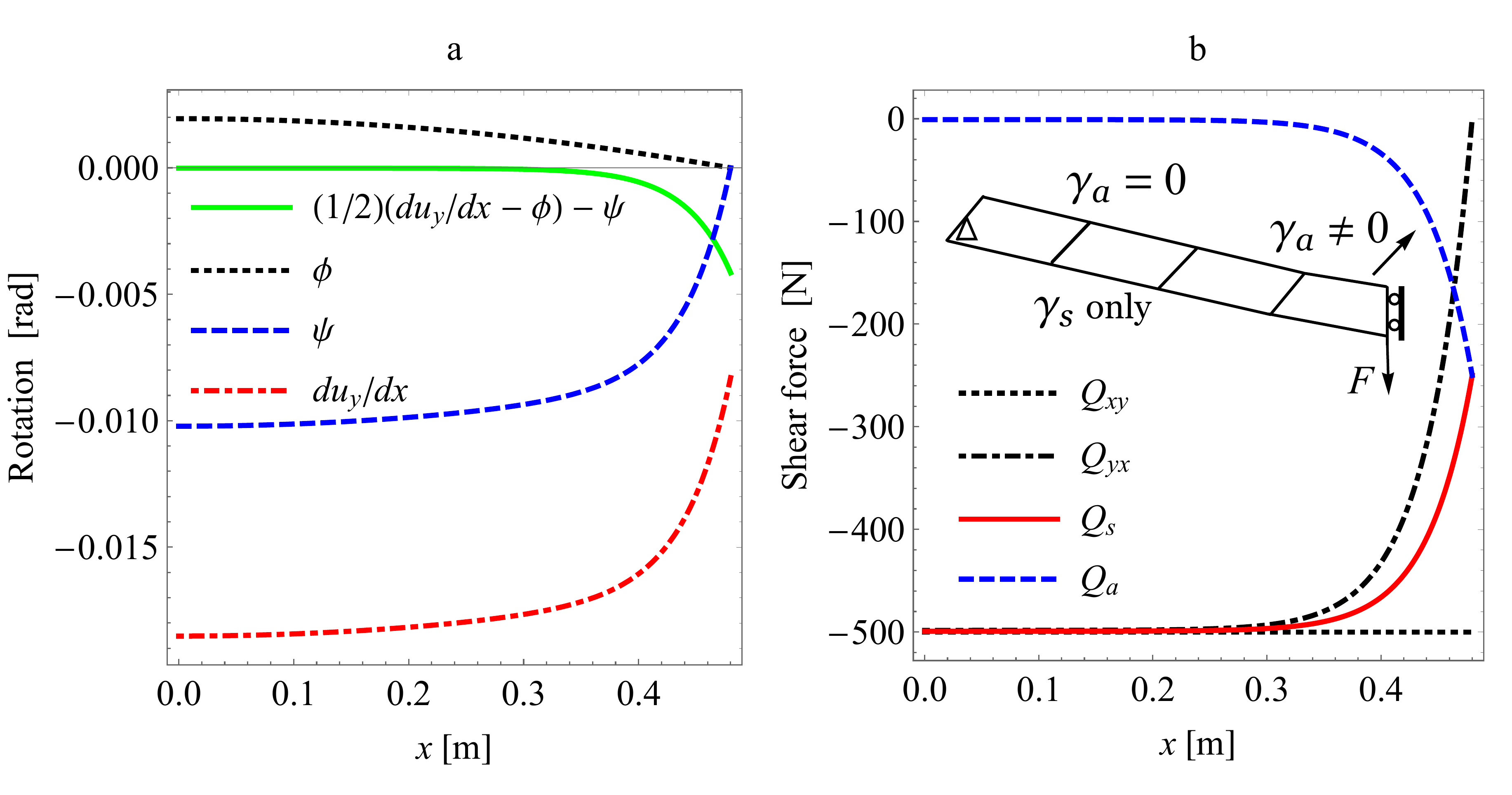}
\caption{(a) Rotation variables of the micropolar ESL Timoshenko beam under three-point bending. The beam consists of four 0.12 m long web-core blocks. (b) Shear forces calculated from the rotations.}
\end{figure}

Figure 11(a) shows the transverse deflection of a four-block web-core cantilever beam under a uniformly distributed load $q_0=1000$ N/m. The boundary conditions are
\begin{equation}
\begin{aligned}
x&=0:\ u_y=\phi=\psi=0, \\ x&=L:\ Q_{xy}=M_x=P_{xz}=0.
\end{aligned}
\end{equation}
In this case, the clamping causes antisymmetric shear deformation of the 1-D micropolar beam in the vicinity of the support. The deflections given by the 1-D classical and couple-stress ESL beams are in error all the way but the results improve when the beam becomes longer. Figure 11(b) shows the moments along the micropolar beam. The total moment, which corresponds to the moment of the classical beam, is given by $M_x-P_{xz}$. We see that there is a small, near-constant moment $P_{xz}$ present also at considerable distance from the clamped boundary. This can be attributed to the distributed load because in the case of a point-loaded cantilever $P_{xz}$ is practically zero after $x=0.12$ m. Note that the couple-stress moment is related to the antisymmetric shear force by
\begin{equation}
\frac{\partial P_{xz}}{\partial x}=-2Q_a,
\end{equation}
according to Eq.~(8) (for $m=0$). Thus, we see from Fig.~11(b) that the antisymmetric shear force is close to zero with distance from the clamped end except for the very tip of the beam when $P_{xz}$ drops to zero and its gradient gives a small antisymmetric shear force.
\begin{figure}[hp]
\centering
\includegraphics[scale=0.74]{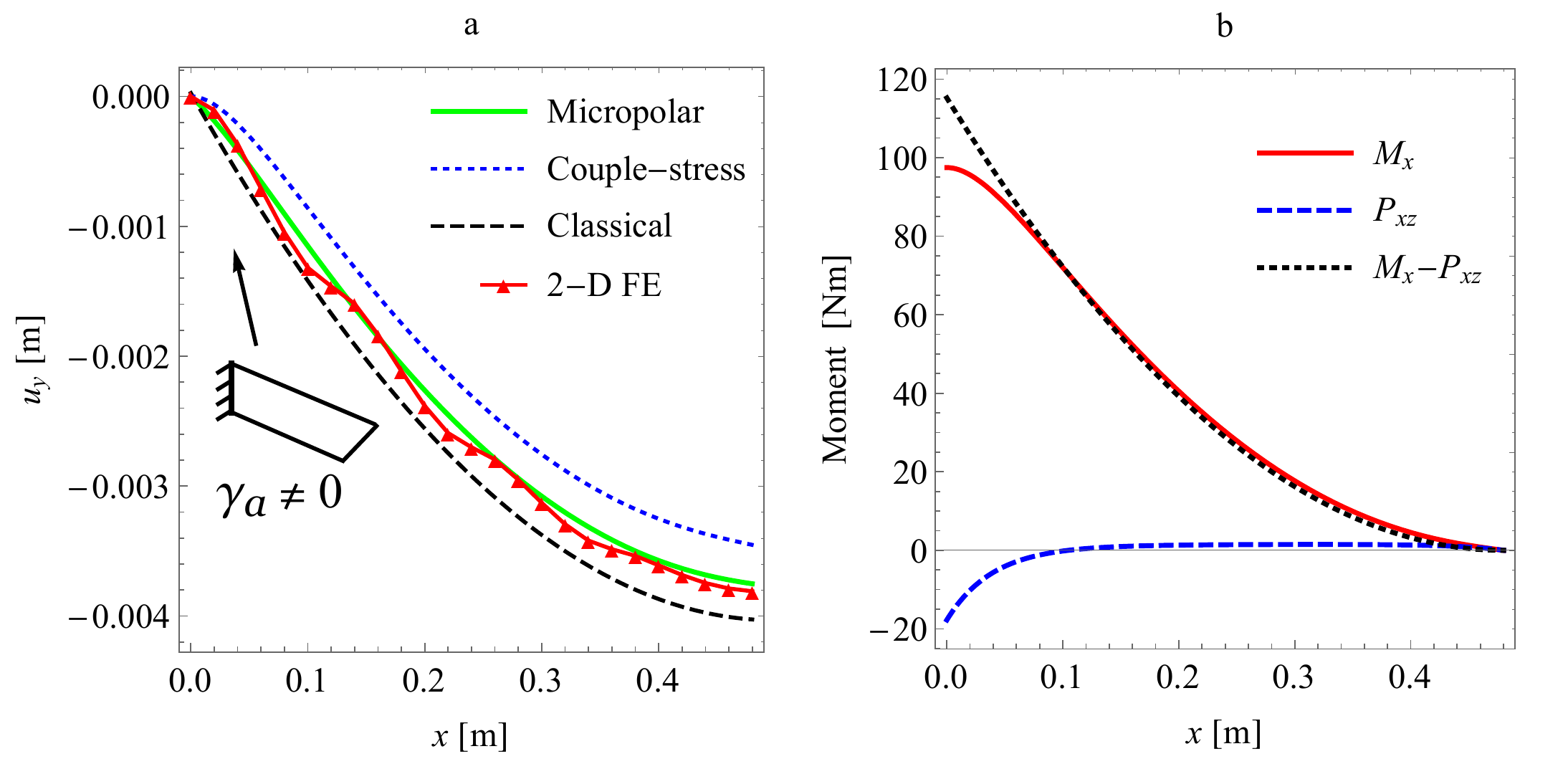}
\caption{(a) Transverse deflections of 1-D classical, couple-stress and micropolar ESL Timoshenko cantilever beams subjected to a uniformly distributed load. (b) Moments along the micropolar beam. The clamping of the shown 2-D web-core block is reflected by notable antisymmetric shear deformation and couple-stress moment $P_{xz}$ in 1-D.}
\end{figure}

\begin{figure}
\centering
\includegraphics[scale=0.65]{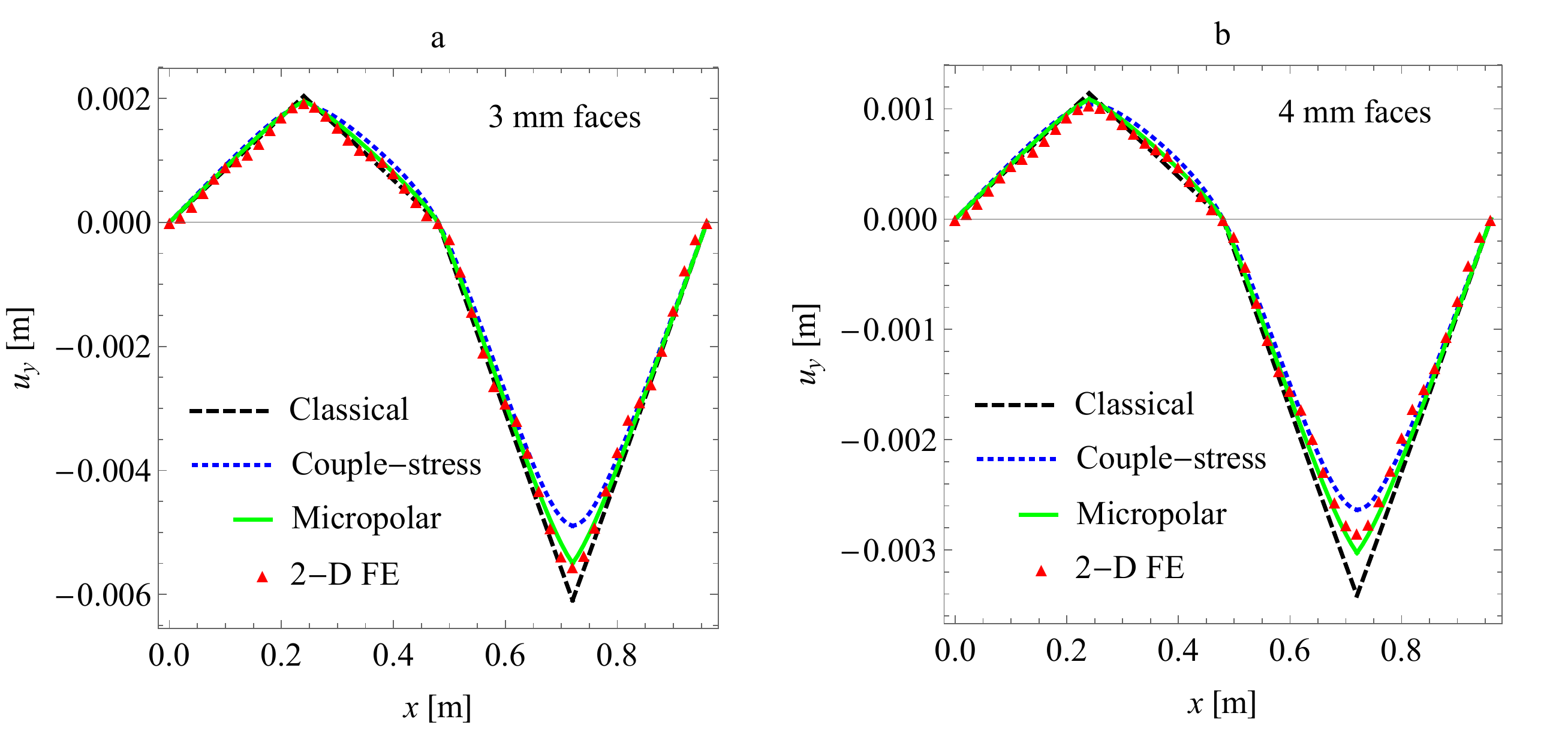}
\caption{Transverse deflection of a beam on multiple supports for (a) 3 mm and (b) 4 mm face sheet thicknesses; see Fig.~8(c).}
\end{figure}
As the last case, we study the beam on three supports presented in Fig.~8(c) to better understand how the micropolar effects may appear in engineering problems modeled in 1-D. For the point loads we have $F=500$ N. Figures 12(a) and 12(b) show the transverse deflection of the beam for 3 mm and 4 mm face thicknesses, respectively. In both cases, the most accurate 1-D response is given by the micropolar ESL Timoshenko beam. It is difficult to determine from Fig.~12 all the places where the micropolar effects are present. Therefore, Fig.~13(a) shows the moments along the beam with 3 mm faces and Fig.~13(b) indicates how the difference between the macrorotation and microrotation manifests along the beam. We see that the micropolar effects appear not only at the locations of the point loads, but also at the mid-support of the beam. The point here is that the couple-stress moment $P_{xz}$ and the difference between the microrotation and macrorotation provide efficient ways to quantify the extent of micropolar effects in engineering problems. In the latter case, the absolute value is used here for better readability because in the ESL model the slope $\partial u_y/\partial x$ of the beam has sharp discontinuities. On a general note, the detailed physical interpretation of the micropolar ESL variables $u_y$, $\phi$ and $\psi$ would require the localization of them so as to obtain the full periodic displacement response of the studied beams. This is left for future studies. Nevertheless, each variable has a clear meaning in an averaged sense in the unit cell analysis (See Fig.~7).
\begin{figure}
\centering
\includegraphics[scale=0.76]{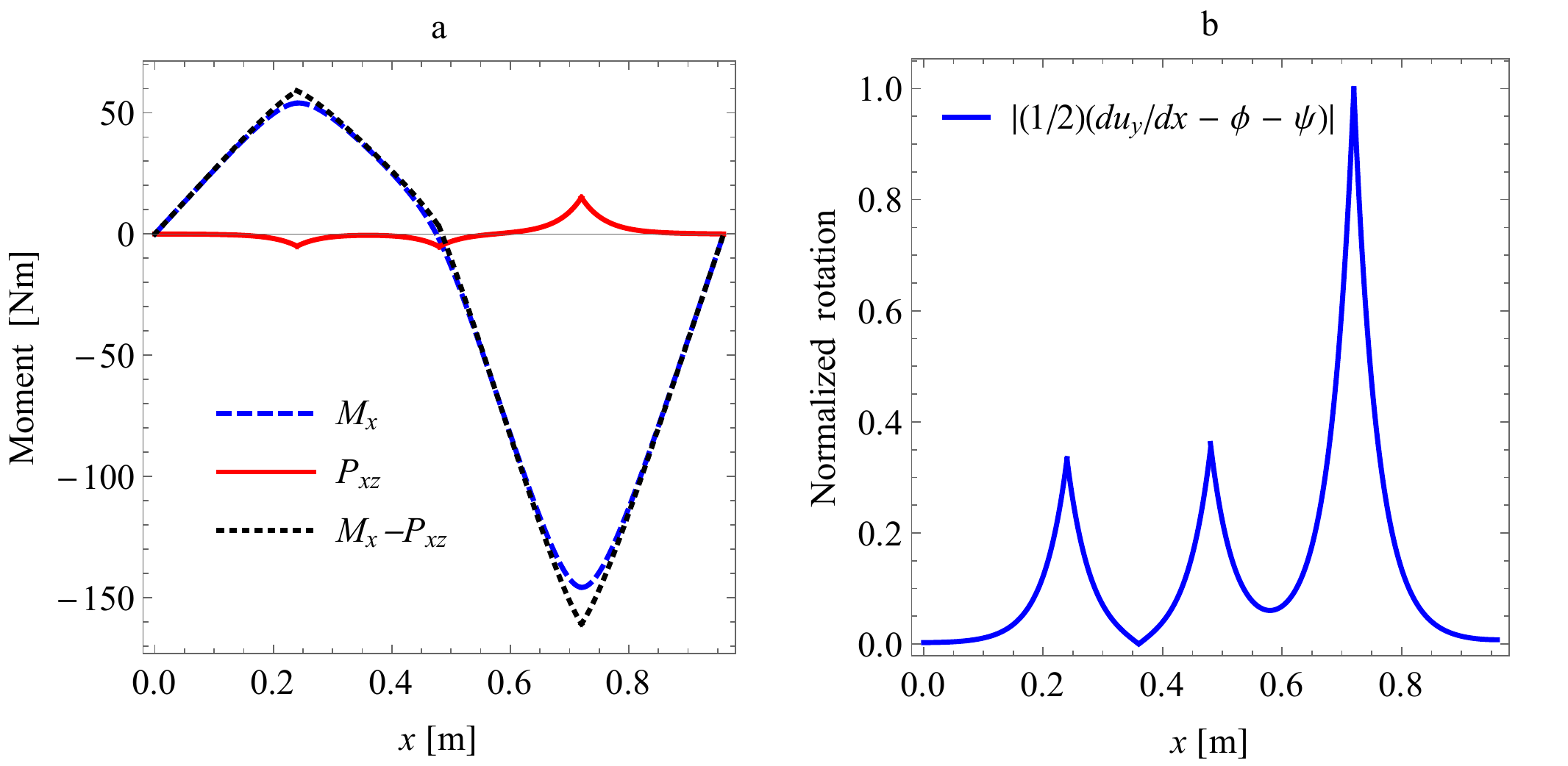}
\caption{(a) Moments along the micropolar beam (3 mm face sheets) on three supports. (b) Absolute value of the difference between macrorotation and microrotation normalized to the maximum value.}
\end{figure}
\section{Conclusions}
In this paper, a micropolar Timoshenko beam theory was formulated, and a nodally-exact finite element using the general displacement solution of the governing beam equations was developed. The beam was used as an equivalent single layer (ESL) model for a web-core sandwich beam. The micropolar ESL bending and shear stiffness coefficients were determined from web-core unit cell analyses. The presented approach may be applied to other core topologies as well.

It was shown that the micropolar ESL Timoshenko beam gives better results in static bending problems than its classical and modified couple-stress counterparts. This is due to the fact that the micropolar approach can capture the antisymmetric shear behavior which occurs in the vicinity of beam supports and point loads. We emphasize that the constituents of a 2-D web-core beam frame do not exhibit any antisymmetric shear strains, but when the 2-D problem is reduced to a 1-D ESL beam problem, the antisymmetric behavior needs to be considered. The antisymmetric shear force is related to the couple-stress moment through equilibrium equations. For a sandwich beam modeled in 1-D this means that the bending of the face sheets by a (non-constant) couple-stress moment causes also antisymmetric shear deformation. The modified couple-stress ESL beam can account for the local face bending; however, it considers only symmetric shear behavior. It follows that the 1-D couple-stress model is too stiff in comparison to 2-D results and the micropolar model, albeit better than the classical ESL model in many respects \cite{goncalves2017}. All in all, the micropolar approach for modeling lightweight sandwich structures has the greatest potential for further development.
\section*{Acknowledgements}
The first author acknowledges that this work has received funding from the European Union's Horizon 2020 research and innovation programme under the Marie Sk\l{}odowska--Curie grant agreement No 745770. The financial support is greatly appreciated.
\appendix
\section{Solution to equilibrium equations}
By differentiating Eq.~(27) once with respect to $x$, we can write
\begin{equation}
D_a(u_y''-\phi'-2\psi')=D_s(u_y''+\phi')-D_x\phi'''.
\end{equation}
Equation (28) then gives
\begin{equation}
u_y''=\frac{D_x}{2D_s}\phi'''-\phi'-q.
\end{equation}
Now, by substituting (A.2) into Eq.~(29) and then differentiating the resulting equation twice with respect to $x$, and noting that according to Eq.~(30)
\begin{equation}
\phi'''=\frac{2D_{xz}}{D_x}\psi'''-\frac{q}{D_x}
\end{equation}
we finally obtain
\begin{equation}
D_{xz}\left(\frac{1}{D_a}+\frac{1}{D_s}\right)\psi'''''-2\left(1+\frac{2D_{xz}}{D_x}\right)\psi'''=\left(1+\frac{1}{2D_s}\right)q''-\frac{2}{D_x}q.
\end{equation}
If we consider only the homogeneous case ($q=0$), the solution to the uncoupled ordinary differential equation (A.4) is given by Eq.~(33). To continue, we introduce two additional variables that have also been used for couple-stress beams \cite{asghari2011,karttunen2016}
\begin{equation}
\gamma\equiv u_y'+\phi \quad \textrm{and} \quad \omega\equiv u_y'-\phi.
\end{equation}
The equilibrium equations (27)--(29) can now be written as
\begin{align}
&\frac{D_x}{2}(\gamma''-\omega'')-D_s\gamma+D_a(\omega-2\psi)=0, \\
&\gamma'=\frac{D_a}{D_s}\left(2\psi'-\omega'-\frac{q}{D_a}\right), \\
&\omega=2\psi-\frac{D_{xz}}{D_a}\psi'',
\end{align}
respectively. Since $\psi$ is known, we obtain $\omega$ directly from (A.8), after which (A.7) yields $\gamma$ through integration. The resulting integration constant $C$ is solved by substituting $\omega$ and $\gamma$ into (A.6), we get (for $q=0$)
\begin{equation}
C=-\frac{2(D_x+D_{xz})}{D_s}c_4.
\end{equation}
Finally, rotation $\phi$ and transverse deflection $u_y$ are retrieved from
\begin{equation}
u_y'=\frac{1}{2}(\gamma+\omega) \quad \textrm{and} \quad \phi=\frac{1}{2}(\gamma-\omega),
\end{equation}
see Eqs.~(31) and (32). The integration of $u_y'$ produces constant $c_1$ which corresponds to a rigid body translation.

\bibliographystyle{model3-num-names}
\bibliography{litera}





\end{document}